\documentclass[twocolumn,showpacs,prb,amsfonts,amsmath,amssymb,floatfix]{revtex4} 
\usepackage{color}
\usepackage{hhline}
\usepackage{mathrsfs}
\usepackage{graphicx}
\usepackage{dcolumn}
\usepackage{bm}
\usepackage{multirow,bigdelim}
\usepackage{booktabs}
\usepackage{ulem}
\input{epsf}
\usepackage{amsmath}

\def\Hscf1{{\cal H}_{SCF}}
\def\dvscf1{\Delta V_{SCF}}
\def\Hscfkq1{{\cal H}_{SCF}^{{\mathbf k}+{\mathbf q}} }
\def\kq1{{\mathbf k} + {\mathbf q}}
\def\k1{{\mathbf k}}
\def\kp1{{\mathbf k}'}
\def\q1{{\mathbf q}}
\def\rp1{{\mathbf r}'}
\def\thetan1{\tilde{\theta}_{F,n}}
\def\thetam1{\tilde{\theta}_{F,m}}
\def\thetamn1{\tilde{\theta}_{m,n}}
\def\thetanm1{\tilde{\theta}_{n,m}}
\def\g1{{\mathbf G}}
\def\gp1{{\mathbf G}'}

\begin{document}

\title{Nonlocal correlations induced by Hund's coupling: a cluster DMFT study}

\author{Yusuke Nomura}
\email{yusuke.nomura@riken.jp}
\affiliation{Department of Applied Physics, University of Tokyo, Hongo, Bunkyo-ku, Tokyo 113-8656, Japan.}
\author{Shiro Sakai}
\author{Ryotaro Arita}
\affiliation{Center for Emergent Matter Science, RIKEN, Hirosawa, Wako, Saitama 351-0198, Japan.}

\date{\today}

\begin{abstract}
We study spatial correlation effects in multiorbital systems, especially in a paramagnetic metallic state subject to Hund's coupling. We apply a cluster extension of the dynamical mean-field theory (DMFT) to the three-orbital Hubbard model away from half filling, where previous single-site DMFT studies revealed that {\it local} correlation effects caused by Hund's coupling bring about unusual strongly-correlated metallic behaviors.
We find that Hund's coupling significantly affects the {\it nonlocal} correlations, too; it strongly modulates the electron distribution in the momentum space so as to make a momentum region almost half-filled and hence strongly correlated.
It leads to an anomalous electronic state distinct both from the Fermi liquid and the Mott insulator. 
We identify the mechanism of the anomalous state with the intersite ferromagnetic correlations induced by Hund's coupling.
\end{abstract} 
\maketitle

\section{Introduction}\label{Sec:intro}

Hund's coupling is known to play essential roles in various electronic properties of strongly correlated materials such as transition-metal oxides. It aligns the electronic spins at each atomic site to maximize the total spin.
In recent years, its substantial role in {\it paramagnetic metals} has also become recognized:~\cite{doi:10.1146/annurev-conmatphys-020911-125045} Haule and Kotliar found, in their study on iron-based superconductors, that it is Hund's coupling $J$ rather than the Hubbard $U$ that induces the 
strong correlation effects in these materials.\cite{1367-2630-11-2-025021} They coined a term ``Hund's metal'' for this state.
For degenerate $N$-orbital systems at integer fillings (except for the fillings $n= 1, N$, and $2N-1$), 
$J$ has more complicated effects: While $J$ increases the effective mass of quasiparticles for small $U$, it allows the quasiparticles to survive up to a rather large $U$ (``Janus-faced" behavior). This behavior has been discussed in relation with diverse metallic properties of perovskite-type transition-metal oxides.\cite{doi:10.1146/annurev-conmatphys-020911-125045,PhysRevB.67.035119,PhysRevLett.107.256401}
More insights were obtained by Werner {\it et al.}, who observed a frozen local-spin moment in a metallic state of the three-orbital Hubbard model away from half filling (``spin-freezing'' behavior).\cite{PhysRevLett.101.166405} 
It was also pointed out that Hund's coupling suppresses orbital correlations (``band decoupler'') and leads to the orbital-selective Mott transition \cite{epjb_Anisimov,PhysRevLett.92.216402} in nondegenerate multiorbital models.\cite{PhysRevLett.102.126401,PhysRevB.83.205112} 

These exotic behaviors have been revealed by the dynamical mean-field theory (DMFT)~\cite{RevModPhys.68.13} applied to the multiorbital Hubbard models. The DMFT studies show that these behaviors originate from the {\it local} correlation effects. 
However, in real materials, {\it nonlocal} correlation effects, which are neglected in the DMFT, may also play essential roles.
To study the latter effects, cluster extensions of the DMFT (cDMFT)~\cite{RevModPhys.77.1027} have been developed.
In fact, the cDMFT applied to the single-orbital Hubbard model has 
revealed that  Hubbard $U$ significantly affects the spatial correlations, inducing a momentum differentiation of the electronic structure especially in a metallic state close to the Mott metal-insulator transition.\cite{RevModPhys.77.1027}

Nevertheless,
 the effect of Hund's coupling on the {\it nonlocal} correlations is an open issue. 
 While there are several cDMFT studies for multiorbital systems in literature,\cite{PhysRevLett.94.026404,PhysRevLett.93.086401,PhysRevB.76.085101} they focused on the effects of e.g., structural distortion and intersite interaction on the nonlocal correlations. In addition, in these studies the spin-flip and pair-hopping interactions are neglected so that Hund's coupling does not preserve the rotational symmetry of spin.
A notable exception is the work by Kita {\it et al.},\cite{PhysRevB.79.245128} who investigated the role of the rotationally-invariant Hund's coupling by the cDMFT combined with the non-crossing approximation and  
found an anomalous metal with a pseudogap in a two-orbital system.\cite{PhysRevB.79.245128} 
However, a study with a more serious treatment of the correlation effects, as well as a study on the three-orbital systems which are relevant to many real transition-metal oxides, are still missing.
This is because the cDMFT with numerically exact solvers has been computationally too expensive to apply to the multiorbital models. 
Especially, if one employs the quantum Monte Carlo (QMC) method to solve the multiorbital impurity problem of the cDMFT, it has been a challenge to preserve the spin-rotational symmetry. Recently, we have developed an efficient sample-update algorithm~\cite{PhysRevB.89.195146} for the continuous-time QMC (CT-QMC) method~\cite{RevModPhys.83.349} based on the interaction expansion.\cite{JETPLett.80.61,PhysRevB.72.035122,PhysRevB.76.035116} The development enabled us the cDMFT calculations for the multiorbital Hubbard models within a reasonable computational time.\cite{PhysRevB.89.195146}

\begin{figure*}[htbp*]
\vspace{0cm}
\begin{center}
\includegraphics[width=0.93\textwidth]{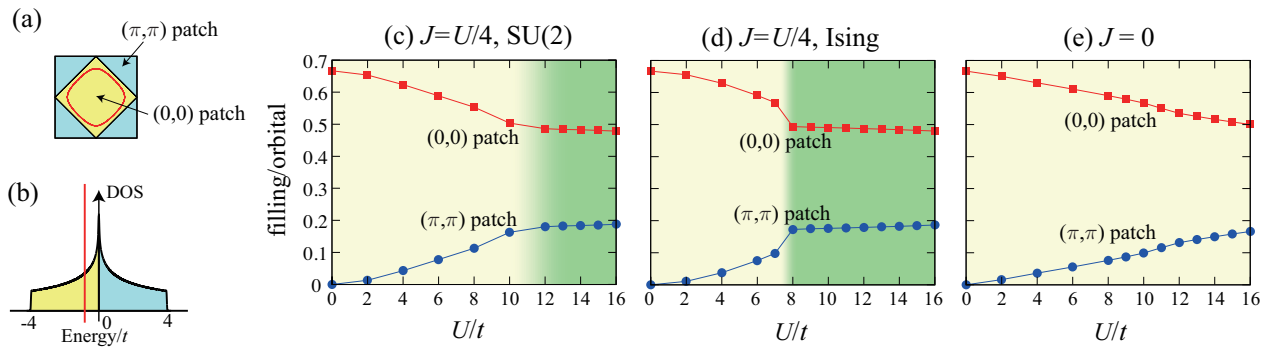}
\caption{(Color online) (a) Momentum patches used in the present DCA calculation. We call the inner [outer] patch $(0,0)$ [$(\pi,\pi)$] patch.
The red curve indicates the Fermi surface in the non-interacting case.  
(b) The non-interacting density of states for each patch. The Fermi level at $U=0$ is shown by a red line. 
(c) [(d)] The DCA results of the fillings in the $(0,0)$ and $(\pi,\pi)$ patches for  $J=U/4$ with [without] the spin-flip and pair-hopping terms. (e) The result for $J=0$.
} 
\label{fig_filling}
\end{center}
\end{figure*}

In this paper, we investigate the spatial correlation effects in the presence of Hund's coupling for
the degenerate three-orbital Hubbard model at the electron filling $n=2$, 
where the previous single-site DMFT studies revealed that Hund's coupling causes the exotic behaviors (such as the ``Janus-faced"\cite{PhysRevLett.107.256401}  and ``spin-freezing"\cite{PhysRevLett.101.166405} behaviors) through the local correlation effects.
With the cDMFT, we find that Hund's coupling induces a drastic filling rearrangement in the momentum space.
It occurs in a way that keeps a momentum region nearly half filled, and thereby makes the region conspicuously strongly correlated, leading to an anomalous metal.
We elucidate that this filling rearrangement is caused by intersite ferromagnetic correlations 
originating from
the interplay between Hund's coupling and electron transfers. 

The paper is organized as follows: 
In Sec.~\ref{sec:model}, we introduce the three-orbital Hubbard model and describe a method to analyze it.
In Sec.~\ref{sec:result}, we show the results for electron distribution in the momentum space, Green's function, self-energy and correlation functions.
The physics behind the unusual momentum differentiation is discussed in Sec.~\ref{sec:discuss}.
Finally,  in Sec.~\ref{sec:summary} we summarize our work and briefly discuss future perspectives.

\section{Model and method}\label{sec:model}
We study the degenerate three-orbital Hubbard model on a square lattice, which is a 
simple model for, for example, layered transition-metal oxides with $t_{2g}$ bands around the Fermi level. 
With this simple model, we aim at extracting a general effect of Hund's coupling, independent of material details.
The Hamiltonian is given by
\begin{eqnarray}
 \mathcal{H}  \ 
&=&  \ 
 -t  \! \! \!  \sum_{\langle i,j \rangle, l ,\sigma}  \! \!    \hat{c}^{\sigma \dagger}_{l i} \hat{c}^{\sigma}_{l j} 
 - \mu    \sum_{i, l, \sigma}   \hat{n}^{\sigma}_{l i} 
+ 
 U \sum_{i,l}  \hat{n}^{\uparrow}_{l i}  \hat{n}^{\downarrow}_{ l i } 
\nonumber \\ 
 &+& \  
U'  \! \!  \sum_{i,l<m,\sigma} \! \! 
 \hat{n}^{\sigma}_{l i}  \hat{n}^{\overline{\sigma}} _{m i}  
+
(U'-J) \! \! \sum_{i, l<m,\sigma} \! \! 
 \hat{n}^{\sigma}_{l i}  \hat{n}^{\sigma}_{m i} 
 \nonumber \\ 
 &+& \ 
 J \!  \sum_{i, l \neq m }  
\hat{c}^{\uparrow \dagger}_{li} 
\hat{c}^{\uparrow}_{mi} 
\hat{c}^{\downarrow \dagger}_{mi}
\hat{c}^{\downarrow}_{li}
+   
J  \!  \sum_{i, l \neq m} 
\hat{c}^{\uparrow \dagger }_{li} 
\hat{c}^{\uparrow}_{mi} 
\hat{c}^{\downarrow \dagger}_{li}
\hat{c}^{\downarrow}_{mi}, \label{hamiltonian}
\end{eqnarray}
where $\hat{c}_{l i}^{\sigma \dagger}$ ($\hat{c}^{\sigma}_{l i}$) creates (annihilates) an $l$th-orbital electron ($l=$ 1-3) with spin $\sigma$ at site $i$ and $\hat{n}^{\sigma}_{l i}\equiv \hat{c}_{l i}^{\sigma \dagger} \hat{c}^{\sigma}_{l i}$. $t$ is a transfer integral between neighboring sites, denoted by $\langle i,j \rangle$. $\mu$ is the chemical potential.
For interaction parameters, we introduce intraorbital Coulomb repulsion $U$, interorbital Coulomb repulsion $U'$, and Hund's coupling $J$ with the relation $U' = U - 2J$. The spin-exchange and pair-hopping terms [3rd line in Eq.~(\ref{hamiltonian})] are necessary to keep the spin-SU(2) symmetry.
Hereafter, we use $t=1$ as a unit of energy. The electron density per site is set to be $n=2$ (i.e., two electrons per site), where it is argued that Hund's coupling brings about the exotic local correlations, as described in Sec.~\ref{Sec:intro}.

In order to investigate the role of Hund's coupling on nonlocal correlations, 
we analyze the model (\ref{hamiltonian}) with the dynamical cluster approximation (DCA) version of the cDMFT,~\cite{RevModPhys.77.1027,PhysRevB.58.R7475,PhysRevB.64.195130} where we employ two momentum patches illustrated in Fig.~\ref{fig_filling}(a).
We call inner [outer] patch $(0,0)$ [$(\pi,\pi)$] patch. 
The DCA approximates the self-energy to be constant within each patch; we call the one for $(0,0)$ [$(\pi,\pi)$] patch $\Sigma_{00}$ [$\Sigma_{\pi\pi}$].
Short-range spatial correlations are incorporated as the difference between $\Sigma_{00}$ and $\Sigma_{\pi\pi}$.
The correlation effects are calculated through mapping the original lattice model (\ref{hamiltonian}) onto a two-site impurity model embedded in self-consistently determined bath sites.
The most time-consuming part of the computation is to solve the two-site impurity model, for which we employ an improved CTQMC method developed in Ref.~\onlinecite{PhysRevB.89.195146}. 
We restrict ourselves to paramagnetic and paraorbital solutions.\cite{note1}
The temperature $T$ is set to be $0.1 t$ and correspondingly the inverse temperature is $\beta = 10/t$. 
We mainly consider the case of $J = U/4$ and, just for comparison, we also show the results for $J=0$.

\section{Results}\label{sec:result}
In Fig.~\ref{fig_filling}(a), we show the Fermi surface for $n=2$ in the non-interacting limit ($U=0$). The corresponding non-interacting density of states and the Fermi level are shown in Fig.~\ref{fig_filling}(b). We see that at $U=0$, all the electrons reside in the $(0,0)$ patch and the $(\pi,\pi)$ patch is completely empty. 
On the other hand, in the atomic limit ($U \rightarrow \infty$), the fillings for $(0,0)$ and $(\pi,\pi)$ patches should agree because the momentum dependence vanishes. 
Although these two limits are trivial, it is nontrivial how they are connected as the interaction $U$ varies.

Figures~\ref{fig_filling}(c-e) show how the filling of each orbital evolves with the interaction strength for
(c) [(d)] $J=U/4$ with [without] the spin-flip and pair-hopping terms, and for (e) $J=0$. 
At $U=0$, the filling $n_{00}$ for the $(0,0)$ patch is $2/3$, and $n_{\pi\pi}$ = 0. 
As the Hubbard $U$ [and accordingly Hund's coupling $J=U/4$ in (c) and (d)] increases, $n_{00}$ ($n_{\pi\pi}$) gradually decreases (increases) in all the cases.
However, there is a remarkable difference between $J>0$ and $J=0$ cases: For $J>0$ the fillings show a ``plateau" (where the fillings have little dependence on $U$) for a range of large interactions, which is colored by green in the figures, while there is no such plateau for $J=0$.
In the plateau region, $n_{00} \sim 0.5$, i.e., the $(0,0)$ patch is nearly half filled. 
Comparing Figs.~\ref{fig_filling}(c) and (d), we find that the spin-flip and pair-hopping terms give a quantitative difference in the results: the density-density-type interaction [Fig.~\ref{fig_filling}(d)] considerably overestimates the tendency toward the filling plateau. 
In the following, we concentrate on the spin-SU(2)-symmetric Hamiltonian and elucidate the physical origin of the plateau behavior. 

\begin{figure}[tbp]
\vspace{0cm}
\begin{center}
\includegraphics[width=0.46\textwidth]{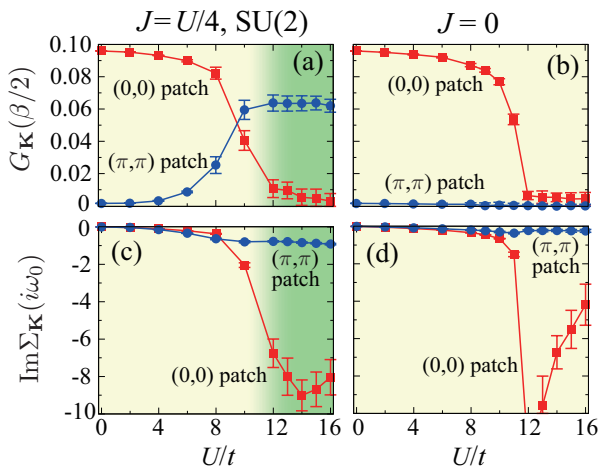}
\caption{(Color online) 
(a),(b) Green's functions at the imaginary time $\tau=\beta/2$, $G_{\bf K}(\beta/2)$, 
and (c),(d) the imaginary parts of the self-energies, ${\rm Im} \Sigma_{\bf K} (i \omega_0)$, at the first Matsubara frequency $\omega_0=\pi T$, for $(0,0)$ and $(\pi,\pi)$ patches. 
The panels (a) and (c) [(b) and (d)] show the results for $J=U/4$ [$J=0$].  
}
\label{fig_gb2}
\end{center}
\end{figure} 

Since the present two-patch DCA does not have enough resolution in the momentum space, we avoid to discuss the shape of the Fermi surface. 
Instead, we plot, in Figs.~\ref{fig_gb2}(a) and \ref{fig_gb2}(b), Green's function of each patch at the imaginary time $\tau  = \beta/2$,  $G_{\bf K}(\beta/2)$, as a measure of the low-energy spectral weight.
In the non-interacting limit ($U=0$), $G_{00}(\beta/2)$ is finite and $G_{\pi\pi}(\beta/2) \sim 0$, since the whole Fermi surface is encompassed in the $(0,0)$ patch [Fig.~\ref{fig_filling}(a)].
As $U$ increases, these low-energy spectral weights evolve differently depending on $J$.
For $J=0$ [Fig.~\ref{fig_gb2}(b)], $G_{\pi\pi}(\beta/2)$ is always nearly zero up to $U=16$ while $G_{00}(\beta/2)$ starts from a finite value at $U=0$, gradually decreases as $U$ increases, and eventually ends up with a very small value at around $U=12$, where the Mott transition occurs.
Consistent with the behavior of $G_{00}(\beta/2)$, $-$Im$\Sigma_{00}(i\omega_0)$, which approximates the scattering rate, shows a rapid increase at around $U=12$ [Fig.~\ref{fig_gb2}(d)].
These results show that the $(\pi,\pi)$ patch never acquires a low-energy spectral weight and that the whole system transits from a Fermi-liquid-like metal to the Mott insulator at a strong interaction ($U\sim 12$).

A remarkable difference is seen in the result for $J=U/4$  [Fig.~\ref{fig_gb2}(a)]. 
In this case, $G_{\pi\pi}(\beta/2)$ acquires a finite weight in a region of $U\gtrsim 6$, where $G_{00}(\beta/2)$ starts to lose its weight. 
For $U\gtrsim 12$, $G_{\pi\pi}(\beta/2)$ dominates the low-energy weight and $G_{00}(\beta/2)$ virtually vanishes, indicating an anomalous electronic state distinct from both the Fermi liquid and the Mott insulator.  As seen in Fig.~\ref{fig_gb2}(c), $-$Im$\Sigma_{\pi\pi}(i\omega_0)$ remains a moderate value ($\sim 1$) in this region. 
Interestingly, this new state of matter emerges in the same parameter region as that of the filling plateau [Fig.~\ref{fig_filling}(a)], as we have indicated by green color. 
Note that the Mott transition in this case is expected to occur at $U>16$, suggesting that the critical value of $U$ is enhanced by $J$.
Although a similar Hund-induced enhancement of the critical $U$ has been seen in the single-site DMFT results,\cite{PhysRevLett.107.256401} 
the enhancement in our case involves the momentum differentiation, which is neglected in the single-site DMFT.

\begin{figure}[htbp]
\vspace{0cm}
\begin{center}
\includegraphics[width=0.35\textwidth]{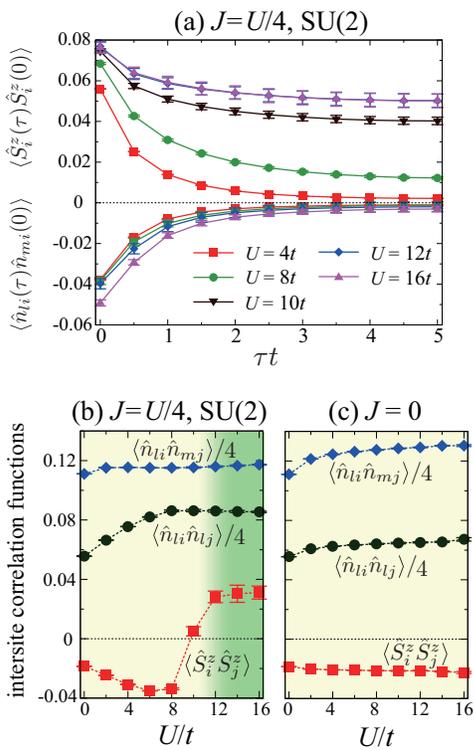}
\caption{(Color online) (a) The onsite spin-spin correlation function $\langle  \hat{S}^{z}_{i} (\tau) \hat{S}^{z}_{i} (0) \rangle$ and orbital-orbital correlation function $\langle  \hat{n}_{l i} (\tau) \hat{n}_{mi} (0) \rangle$ for $J=U/4$. The spin-spin and orbital-orbital correlation functions take positive and negative values, respectively.
(b),(c) The intersite spin-spin correlation function
 $\langle  \hat{S}^{z}_{i}  \hat{S}^{z}_{j}  \rangle$, 
 intraorbital density-density correlation function
$\langle  \hat{n}_{l i}  \hat{n}_{lj}  \rangle/4$,
and interorbital density-density correlation function
$\langle  \hat{n}_{l i}  \hat{n}_{mj}  \rangle/4$ 
for $J=U/4$ and $J=0$, respectively. } 
\label{fig_cor}
\end{center}
\end{figure} 

Figure \ref{fig_cor}(a) shows the onsite spin-spin correlation function $\langle  \hat{S}^{z}_{i} (\tau) \hat{S}^{z}_{i} (0) \rangle$ and orbital-orbital correlation function $\langle  \hat{n}_{l i} (\tau) \hat{n}_{mi} (0) \rangle$ for $J=U/4$. 
Here, $S^{z}_{i} = \frac{1}{3}\sum_{l} \frac{1}{2} ( \hat{d}^{\uparrow \dagger}_{li} \hat{d}^{\uparrow}_{li} - \hat{d}^{\downarrow \dagger}_{li} \hat{d}^{\downarrow}_{li}) $ and $\hat{n}_{l i}  = \sum_{\sigma} \hat{d}^{\sigma \dagger}_{li} \hat{d}^{\sigma}_{li} $ with 
$\hat{d}_{l i}^{\sigma \dagger}$ ($\hat{d}^{\sigma}_{l i}$) being a creation (an annihilation) operator of electrons in the $l$th-orbital ($l=$ 1-3) with spin $\sigma$ at the impurity site $i$.
We see that the orbital-orbital correlation function $\langle \hat{n}_{l i}(\tau) \hat{n}_{mi} (0) \rangle$ is negative and its amplitude decays exponentially with $\tau$.
On the other hand, the spin-spin correlation function $\langle  \hat{S}^{z}_{i} (\tau) \hat{S}^{z}_{i} (0) \rangle$ is positive and remains substantially large up to $\tau = \beta/2$ when $U$ and $J$ are large, indicating the presence of a frozen moment (i.e., finite component at $\omega = 0$).
The qualitative behaviors of these local correlation functions are consistent with those obtained with the single-site DMFT.\cite{PhysRevLett.101.166405}

Next, let us turn to the nonlocal correlations of our main interest.
Figures \ref{fig_cor}(b) and \ref{fig_cor}(c) show the intersite correlation functions at $\tau =0$; the spin-spin channel,
 $\langle  \hat{S}^{z}_{i}  \hat{S}^{z}_{j}  \rangle$, 
intraorbital density-density channel,
$\langle  \hat{n}_{l i}  \hat{n}_{lj}  \rangle/4$,
and interorbital density-density channel,
$\langle  \hat{n}_{l i}  \hat{n}_{mj}  \rangle/4$.
In the absence of Hund's coupling ($J=0$), the intersite spin-spin correlation is always antiferromagnetic (negative) in the range from $U=0$ to $U=16$.
However, for $J=U/4$, the intersite spin-spin correlation changes from antiferromagnetic to ferromagnetic around $U=10$. 
The ferromagnetic region in Fig.~\ref{fig_cor}(b) agrees well with the region showing the filling plateau in Fig.~\ref{fig_filling}(c). 
Thus, in the filling-plateau region, the $(0,0)$ patch [$(\pi,\pi)$ patch] loses [gains] the low-energy spectral weight, and the intersite spin-spin correlation becomes ferromagnetic. 

\section{Discussions} \label{sec:discuss}
\begin{figure}[htbp]
\vspace{0cm}
\begin{center}
\includegraphics[width=0.38\textwidth]{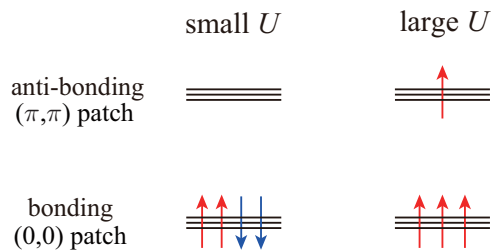}
\caption{(Color online) The momentum-space configurations of four electrons in the two-site cluster with three orbitals at each site. 
The up and down arrows indicate the up- and down-spin electrons, respectively. 
See the main text for more details.} 
\label{fig_config}
\end{center}
\end{figure}

In order to understand the mechanism of these anomalous behaviors, we consider spin-orbital configurations of four electrons in a two-site cluster (with three orbitals at each site) in two different ranges of interaction.
First, in the non-interacting limit ($U=0$), the three orbitals at each site form intersite bonding and antibonding orbitals in the two-site cluster. 
The bonding and antibonding orbitals correspond to $(0,0)$ and $(\pi,\pi)$ components, respectively, in the two-patch DCA.\cite{0295-5075-85-5-57009}
Here, the electron hopping between the two sites lifts the degeneracy of the bonding and antibonding orbitals. This energy offset can be regarded as an effective crystal-field splitting between the molecular orbitals. 
Then, in the ground state all the four electrons occupy the bonding orbitals (Fig.~\ref{fig_config}, small $U$).

Next, we consider the large interaction case.
A simple consideration on the second-order processes of the electron hopping suggests that Hund's coupling makes electrons at neighboring sites align their spins and occupy different orbitals ($\langle  \hat{n}_{l i}  \hat{n}_{mj}  \rangle > \langle  \hat{n}_{l i}  \hat{n}_{lj}  \rangle$).
This ferromagnetic instability leads to a high-spin state, illustrated in the right panel of Fig.~\ref{fig_config}, where the direction of the spin fluctuates dynamically.~\cite{note2}
Note that the above second-order processes can be captured within the single-site DMFT,\cite{Karsten_ferro,PhysRevB.58.R567,PhysRevLett.99.216402,PhysRevB.80.235114} while our cDMFT takes into account the spatial correlations more accurately.
As is seen in the right panel of Fig.~\ref{fig_config}, 
in such a high-spin state, an electron is pumped up to the
antibonding ($\pi,\pi$) orbitals against the crystal-field splitting,
 and the bonding (0,0) orbitals become nearly half filled.
Then, the $(0,0)$ patch would develop strong correlations, as is indeed seen in the loss of the low-energy spectral weight [Fig.~\ref{fig_gb2}(a)] as well as the enhancement of the scattering rate [Fig.~\ref{fig_gb2}(c)].
The strong correlation in the $(0,0)$ patch in turn fixes the electron filling of this patch, 
explaining the filling plateau of the $(0,0)$ patch seen in Fig.~\ref{fig_filling}(c).

\section{Conclusion and outlook}\label{sec:summary}

We have explored spatial correlation effects in multiorbital systems by means of the two-patch DCA. 
In the degenerate three-orbital Hubbard model at $n=2$, we have found that Hund's coupling, in combination with the electron hopping, brings about a drastic rearrangement of the electron distribution in the momentum space.
 In particular, the $(0,0)$ patch is almost fixed at half filling and becomes strongly correlated. 
There the low-energy spectral weight emerges in the $(\pi,\pi)$ sector, indicating that a new state of matter appears in between the Fermi-liquid metal and the Mott insulator.
By calculating the two-particle correlation functions, we have identified the mechanism of this anomalous behavior with the intersite ferromagnetic correlations enhanced by Hund's coupling.
We have found similar behaviors also in the two-orbital Hubbard model at the filling $n=1.5$, which suggests that these exotic behaviors are generic in multiorbital systems away from half filling.


Although larger clusters are intractable at this moment, the competition between the interactions and the effective crystal-field splitting in the momentum space always exists no matter how large the cluster is.
Since the effective crystal-field splitting between the neighboring energy levels becomes smaller as the cluster size increases (and eventually the energy levels become continuous in the thermodynamic limit), it would be natural to expect that Hund's coupling, whose Fourier transform gives intermomentum interactions, plays a crucial role in the competition.
Provided that Hund's coupling makes some patches high-spin state and nearly half-filled, 
it will lead to a redistribution of the electrons in the momentum space. 
The high-spin patches will develop a strong correlation and thereby cause a momentum differentiation. 
Especially, if the high-spin state is realized in patches with the Fermi surface, 
the patches might lose its low-energy spectral weight due to the strong correlations. 
This momentum-dependent loss of the low-energy spectral weight may give a pseudogap behavior in multiorbital systems, caused by Hund's coupling and being distinct from the pseudogap mechanism discussed in single-orbital models.\cite{RevModPhys.77.1027}

Recently, the pseudogap behaviors have been observed in several multiorbital materials: nickelates,\cite{PhysRevLett.106.027001} ruthenates,\cite{PhysRevB.58.R13318,PhysRevLett.112.206403} iridates,\cite{Kim11072014} and the iron-based superconductors.\cite{ncomms1394,PhysRevLett.111.217002,PhysRevLett.109.037002,PhysRevLett.109.027006,PhysRevB.89.045101} 
It is an interesting future issue to study these materials by combining the present scheme with realistic band-structure calculations.

\begin{acknowledgements}
We would like to thank Massimo Capone and Philipp Werner for fruitful discussions. 
Y. N. is supported by Grant-in-Aid for JSPS Fellows (No. 12J08652), 
and S. S. by Grant-in-Aid for Scientific Research (No. 26800179).
The calculations were performed at the Supercomputer Center, ISSP, University of Tokyo, 
Research Institute for Information Technology, Kyushu University, 
and 
 Supercomputing Division, Information Technology Center, University of Tokyo.  
\end{acknowledgements}

\bibliographystyle{apsrev}
\bibliography{PRB}

\begin{thebibliography}{38}
\bibitem[{not({\natexlab{a}})}]{note1}
\bibinfo{note}{The self-consistent loops are started from a metallic solution.}
\bibitem[{not({\natexlab{b}})}]{note2}
\bibinfo{note}{Since we consider the paramagnetic solution, the configuration
  with three down-spin electrons in the bonding orbitals and one down-spin
  electron in anti-bonding orbitals is equally realized in the Monte Carlo
  sampling. Note that configurations in which two or more electrons occupy the
  anti-bonding orbitals also have a small weight.}
\end{thebibliography}

\end{document}